\documentclass[sd,amsmath,amssymb,groupedaddress,reprint]{revtex4-1}
 \usepackage[pdftex]{graphicx}
\usepackage{dcolumn}% Align table columns on decimal point
\usepackage{bm}% bold math
\usepackage[table,xcdraw]{xcolor}
\usepackage{float}
\usepackage{blindtext}
\def\be{\begin{equation}}
\def\ee{\end{equation}}

\begin{document}
%\lipsum
%\kant[1-2]
\preprint{AIP/123-QED}

\title{An Epidemic Model SIPHERD and its application for prediction of the spread of COVID-19 infection for India and USA}

\date{\today}

\begin{abstract}
 We propose an epidemic model SIPHERD in which three categories of infection carriers Symptomatic, Purely Asymptomatic, and Exposed are considered with different rates of transmission of infection that are taken dependent on the lockdown and social distancing. The rate of detection of the infected carriers is taken dependent on the tests done per day. The model is applied for the COVID outbreak in Germany and South Korea to validate its predictive capabilities and then applied to India and the United States for the prediction of its spread with different lockdown situations and testing in the coming months.
% The rate of infection is model in terms of lockdown parameter, temperature and population of the country. 
\end{abstract}
%\thanks{ashutosh.mahajan@vit.ac.in}

\author{Ashutosh Mahajan}
 \affiliation{Centre for Nanotechnology Research, Vellore Institute of Technology, Vellore-632 014, India}
 \email{ashutosh.mahajan@vit.ac.in}
\author{Ravi Solanki}
 \affiliation{Centre for VLSI and Nanotechnology, Visvesvaraya National Institute of Technology, Nagpur-440 010, India}
 \author{Namitha A. S.}
  \affiliation{School of Electronics Engineering, Vellore Institute of Technology, Vellore-632 014, India}
\date{\today}

\maketitle

\begin{section}{ Introduction}

The outbreak of pandemic Coronavirus disease 2019 (COVID-19) has led to more than 4 million total infections and 285 thousand  deaths worldwide \cite{worldo}, and serious efforts are needed for its containment. The Coronavirus SARS-CoV-2 has affected not just the public health but made a drastic impact on the economy of the world as well due to the lockdown situations in many countries.

Pandemics have hit humanity many times in the past as well, and mathematical models are already available for infectious diseases. Modeling and simulation can help to predict the extent of the contagious disease and can give useful inputs on correction measures for its containment. In order to devise the lockdown strategy, it is important that the prediction of the disease spread is available to the decision-makers. COVID-19 is different from the previously known SARS (Severe acute respiratory syndrome)  infection, such as the existence of purely asymptomatic cases \cite{diamond} and the spread of the infection from them as well as from the exposed ones in the incubation period \cite{german}. Our proposed mathematical model, SIPHERD  incorporates the above facts for the COVID-19 epidemic.

Many epidemiological models exist in the literature, and the  basic SIR model \cite{SIR} is the  widely used one, which needs to be modified to incorporate the complexities involved in Coronavirus spread and control. An  approximate mathematical model of the COVID-19 is initially reported in the literature \cite{ref4}  based on the  Between-Countries Disease Spread (Be-CoDiS), which  is a spatial epidemiological  model for the study of the spread of human diseases between and within the  countries. 

An improved  mathematical model for the spread of  COVID-19 is proposed in \cite{ref5},  by taking into account the infected and undetected cases. But  this study and forecast is  particularly based only on China.  

An extended SIR model is proposed in \cite{sidarthe}, in which the entire people in the country are divided into eight compartments. Though it is an improved version of the SIR model, the study and simulation results are done only for Italy, and the model does not take into account purely Asymptomatic cases and the role of tests done per day. Another  compartmental epidemic model SEIR  \cite{ref8} forecast for few countries and the impact of the quarantine on the COVID-19 is investigated. A better adaptive and improved version of the SIR model is illustrated in \cite{ref9}. In this method, the  time dependency of some parameters used for the analysis  makes it more robust than the conventional SIR method.
Some other  curve fitting based methods are also available in the literature  for the forecast of COVID-19  in \cite{ref10}, \cite{ref11} and \cite{ref12}. Although these methods can track the  available data correctly, they are not developed based on the  physical insights that  affect the rate of spreading  of the disease and also it is extremely sensitive to the initial conditions.

In this paper, we formulate the mathematical model SIPHERD for the COVID-19 epidemic and apply it for forecasting the number of active cases, confirmed cases, daily new cases, and deaths for India and USA, depending  on the lockdown strategy and the  number of  tests performed per day.

\end{section}

\begin{section}{Model}

We model the evolution of the COVID-19 disease  by dividing  the population into  different categories  as listed below. As seen in FIG. \ref{model}, the rates of transfer  from  one  category to  another can are the model parameters and a set of differential equations for the entity in each category can be formed. We write the model equations that are independent of the population of the country by considering the fraction of the people in each category. 

\begin{itemize}
\item{S} - fraction of the total population that is healthy and has never caught the infection
\item{E} - fraction of the total population  that is exposed to infection, transmit the infection and turn into either Symptomatic or purely Asymptomatic, and not detected 
\item{I} - fraction of the total population infected by the virus that shows symptoms and undetected
\item{P} - fraction of the total population infected by the virus that doesn't show symptoms even after the incubation period and undetected. These are the  purely Asymptomatic cases 
\item{H} - fraction of the total population  that are found positive in the test and either hospitalized or quarantined
\item{R} - fraction of the total population  that has recovered from the infection 
\item{D} - fraction of the total population  that are extinct due to the infection.
\end{itemize}

The SIPHERD model equations are a set of coupled ordinary differential equations (\ref{eq:S} to \ref{eq:D}) for the   defined entities (S,I,P,H,E,R,D), where initial conditions E(0), P(0) and I(0) are not exactly known. The various rates listed in TABLE \ref{para}  are the parameters of the problem  which are also  not known, and only possible range is available. Some of the parameters such as rates of infection ($\alpha$, $\beta$, $\gamma$) change with time in steps, depending on the conditions such as lockdown. 

\begin{figure}[t]
    \includegraphics[width=8.5cm]{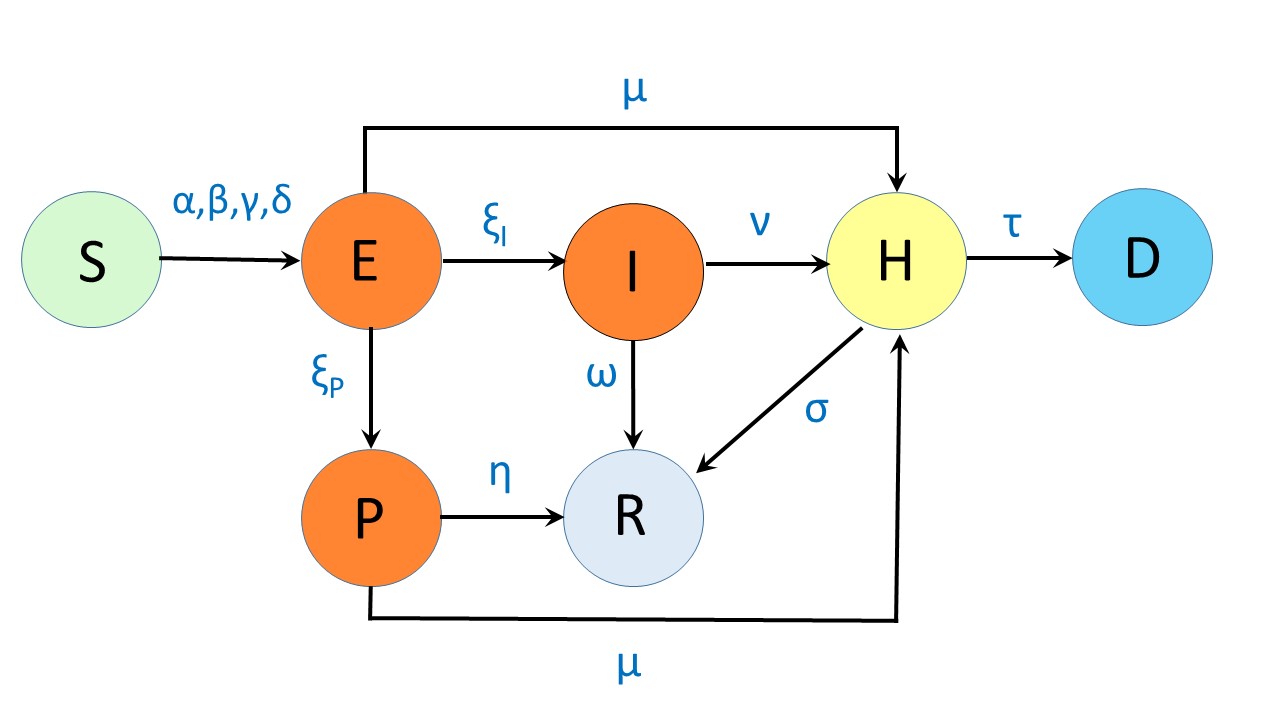}\par\caption{SIPHERD Model} \label{model}
    \end{figure}

\begin{eqnarray}
\frac{dS}{dt}\, \, &=&-S(\alpha E+\beta I +\gamma P+ \delta H) \label{eq:S}\\
\frac{dE}{dt} &=& \, S(\alpha E+\beta I +\gamma P+ \delta H)- (\mu+\xi_I+\xi_P)E\label{eq:E}\\
\frac{dH}{dt}&=&\mu (E+P)+\nu I -\sigma H(t-t_R)-\tau H(t-t_D) \label{eq:H}\\
\frac{dI}{dt}&=& \xi_I E-(\nu+\omega)I \label{eq:I}\\
\frac{dP}{dt}&=& \xi_P E-(\mu +\eta)P \label{eq:P}\\
\frac{dR}{dt}&=&\omega I+\eta P+\sigma H(t-t_R) \label{eq:R}\\
\frac{dD}{dt}&=& \tau H(t-t_D)\label{eq:D}
\label{model_eqn}
\end{eqnarray}
where, $t_R$ and $t_D$ are the delay associated with the recovery and death respectively with respect to active cases $H$. We have taken into account this delay because the active cases are reported after the testing and admission to healthcare or quarantine center and  the number of recovery and death of the admitted will not immediately follow  the  active or $H$ category number. 
All fractions add up to unity that can also be seen from summing the above equations.
\begin{equation}
\frac{d}{dt}( S+I+P+H+E+R+D)=0
\end{equation}

\begin{table}[H]
\caption{}
\begin{tabular}{|l|l|l|l|}
\hline
Parameter & Description   \\ \hline
      $\alpha$    &  Rate of transmission of infection from E to S                        \\ \hline
      $\beta$ &   Rate of transmission of infection from I to S                  \\ \hline
      $\gamma$    &  Rate of transmission of infection from P to S                        \\ \hline
      $\delta$    &  Rate of transmission of infection from H to S                        \\ \hline
      $\xi_{I,P}$ & Rate of conversion from the E to I,P                               \\ \hline
      $\omega$  &     Rate of home recovery of H                         \\ \hline
      $\eta$   &   Rate of home recovery of P               \\ \hline  
      $\mu$  &    Probability of E and P being detected              \\ \hline
      $\nu$   &    Probability of I being detected          \\ \hline
\end{tabular}
 \label{para}
\end{table}

The probability of getting the infection is assumed uniform among the susceptible people, although the disease spreads localised   in hot-spots.

The asymptomatic proportion of the infected persons onboard the Diamond Princess cruise ship is estimated in \cite{diamond}. Among the 634 tested positive onboard,  328 were found asymptomatic i.e., more than 50 percent of the confirmed cases were not showing any specific symptoms of COVID-19. This factor is incorporated in the model by considering the purely Asymptomatic category. The ratio of purely Asymptomatic (P) to total Asymptomatic (E+P) cases is reported to be 0.35 and the ratio of purely Asymptomatic to the total infected (E+P+I) is 0.179 \cite{diamond}. These reported numbers are used to fix the proportion between $\xi_P$ and $\xi_I$ as 0.36 and the proportion of initial conditions $E(0)$, $I(0)$ and $P(0)$ as well. In other words,  out of 100 exposed cases, after the incubation period, 36 will turn to be purely asymptomatic, and 74 will have symptoms.

 Coronavirus-nCoV2 has shown particular characteristics that the asymptomatic patients do transmit the disease. The infection can be transmitted from the person who is not showing illness during the incubation period \cite{german}. This can be included in the model by considering $E$ category people and their transmission as well. Hospitalized and quarantined cases can also transmit the disease, and this small rate is taken as  parameter $\delta$.

The detection of the Asymptomatic and Symptomatic cases can be taken dependent on the number of tests done per day ($ T_{PD}$).
 For the Symptomatic cases, the detection is more probable as the infected person can approach for the tests and more likely to be tested. The detection of Symptomatic is taken in two parts, one a constant and another part proportional to the tests done per day. This can be written in terms of parameters as, 
\be
\nu= \nu_0+\nu_1 T_{PD}
\ee
\be
\mu=\mu_0 T_{PD}
\ee
where, $\mu_0$, $\nu_0$, and $\nu_1$ are positive constants.
Recovery of Asymptomatic cases is taken faster than the Symptomatic cases. The total confirmed cases are the addition of the active cases, extinct cases, and a part of the recovered that were detected. This can be written as
\be
T_c(t)=H (t)+D(t)+\int^{t} \sigma H  (\tau')  d\tau' \ee

%The rate of infection is taken to be dependent on the population density and  temperature in that country
%\be
%\alpha=\alpha_0 (100-L)/100 (50-T)/50
%\ee

\end{section}
\begin{section}{Numerical Implementation and Simulation}
The set of coupled ordinary differential equations for the model can be readily solved numerically for  a given set of parameters and initial values. The non-trivial part is the accurate determination of the parameters that will mimic the situation on the ground. The mathematical problem is to take into account the four actual data sets of the total number of confirmed cases, active cases on a particular day, cumulative  deaths and tests done per day, and  find the set of parameters that will provide the best possible match between the data and model. The extraction of the parameters is also to be automated so that the model can be run on data for various countries. % For this problem, we used Tikhonov Regularization scheme \cite{tikhonov}, where the standard deviation between the actual data and model values can be kept in  "Fidelity" term and other physical imposition such as the ratio of Asymptomatic to Symptomatic cases can be kept in "Penalty"term. A balancing equation can be written between these two terms to find the Tikhonov regularization parameter. 
A cost function is  written in terms of errors between the actual and solver data sets. The minimizer of the cost can be found to obtain the optimized set of parameters that best fit with the data available till date. The model and the optimization scheme is implemented in MATLAB. 

\end{section}

\begin{section}{Results and Discussion}

 We collected the number of total positive or confirmed cases, present active cases and deaths from \cite{worldo}, \cite{covid19india}, and the number of tests per day from \cite{worldo1}, which is plotted in FIG. \ref{tpd}. The day on which lockdown is imposed in a country is also taken into account as changes in the slopes of the data are observed according to it. The rate of transmission of infection from the Asymptomatic carrier ($\alpha$, $\gamma$) for a country is taken higher than the Symptomatic ones  ($\beta$) as the Asymptomatic carrier may not be aware of his/her infection, and  Susceptible may not be keeping distance as no symptoms are seen. The mortality rate ($\tau$) is taken different for different countries as it depends on the immunity and how effectively the critical patients are taken care of by the hospitals. 
 
 \begin{figure}[t]

    \includegraphics[width=\linewidth]{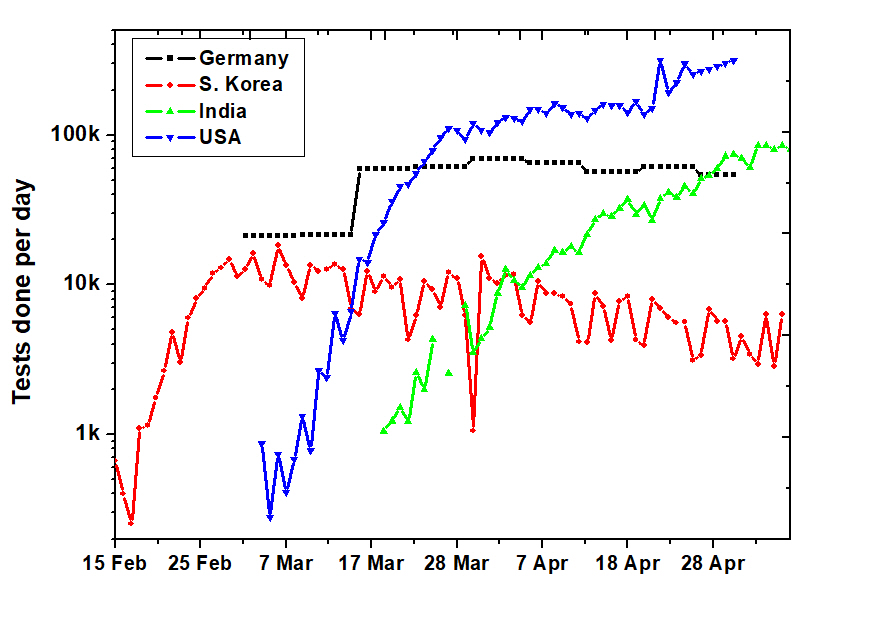}\par\caption{Daily Tests per day data for different countries obtained from \cite{worldo1}} \label{tpd}
 \end{figure}

 The home recovery rates of Asymptomatic ($\eta$) and Symptomatic ($\omega$), rate of transmission of infection from hospitalized and quarantined ($\delta$), and rate of self-reporting of the Symptomatic people ($\nu_0$) are taken uniform for all countries. %We have taken the net number of the purely Asymptomatic and Symptomatic infected originating from the Exposed are taken in 0.36:1 proportion, which is done by taking The transfer rates $\xi_I$ and $\xi_P$ are taken in the proportion of 1:0.36.
 The transfer rate $\xi_I$ is the inverse of the incubation period, whose mean is reported 5.2 days \cite{zhang_lancet}. The parameters determined by our model are listed in TABLE \ref{countries_para} for the countries we studied.  
  
 \begin{figure}[h]

    \includegraphics[width=\linewidth]{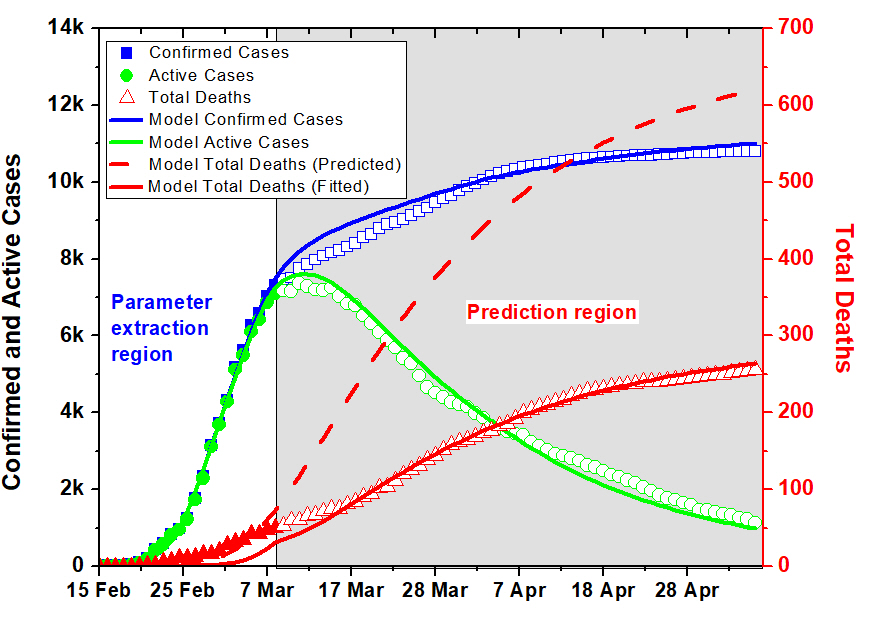}\par\caption{Model prediction using the South Korea data up-to March 9  and comparison with the actual data for the confirmed, active cases and total deaths. } \label{pred_SK}
   \end{figure}
 \subsection{Model Validation}
 
 We apply the SIPHERD model to South Korea and Germany for exhibiting the predictive capability of our model as the disease has almost reached the end stages in these countries.  We used the data for the first 23 and 40 days, respectively, for these counties i.e., till March $9^{th}$ and  March $30^{th}$ and compared the future evolution generated by the model to the actual data as shown in the grey region in FIG. \ref{pred_SK} for South Korea and in FIG.\ref{pred_Germany} for Germany. Model predicted higher deaths for South Korea as the mortality rate 2.2E-3 was higher before March 9 that improved later to fitted value 0.9E-3.

 %We apply the SIPHERD model for South Korea and Germany and extract the parameter values. In South Korea and Germany, the disease has almost reached the end stages, therefore, we used the data for the first 23 and 40 days respectively  i.e., till March $9^{th}$ and  March $30^{th}  $  for checking the predictive power of the model.
 %The comparison of the model-generated data with the actual data after 23 and 40 days is shown in the grey region in FIG. \ref{pred_SK} for South Korea and in FIG. \ref{pred_Germany}.
  \begin{figure}[H]
    \includegraphics[width=\linewidth]{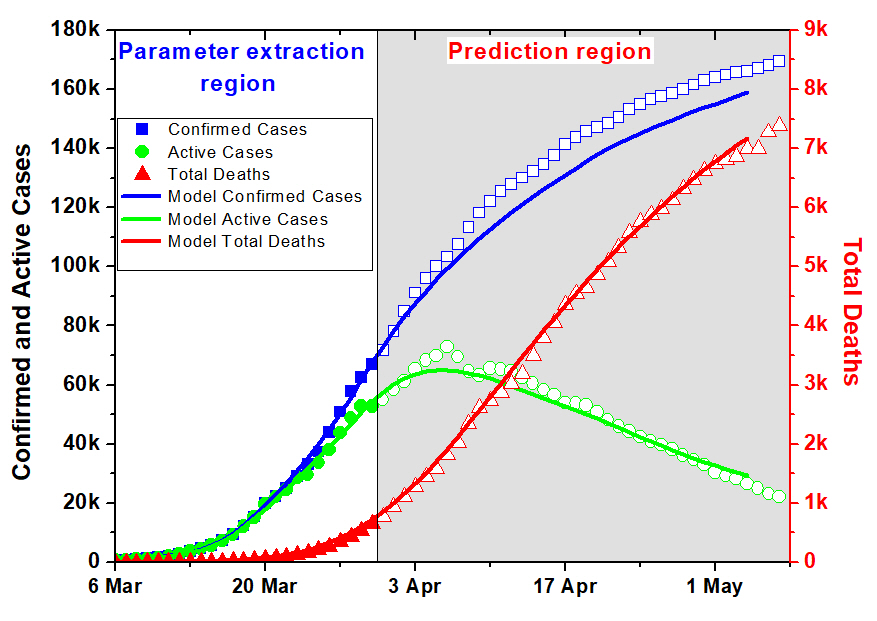}\par\caption{Model prediction using the Germany data up-to March 30 and comparison with the actual data for the confirmed, active cases and total deaths.} \label{pred_Germany}
   \end{figure}
   %  \begin{figure}[H]
    %\includegraphics[width=\linewidth]{plots/new/Germany2.jpg}\par\caption{Model prediction using the Germany data up-to March 30 and comparison with the actual data for the daily new cases. }\label{pred_Germany1}
    %\end{figure}
  %\begin{figure}[h]
   % \includegraphics[width=\linewidth]{plots/new/Germany3.jpg}\par\caption{Number of Infected in Exposed (E), Symptomatic (I) and purely Asymptomatic (P) category for Germany. }\label{pred_Germany2}
    %\end{figure}
   \subsection{Predictions for India}
 For the available data till date, we run the model to extract parameters, and then with the extracted parameters, the  model is run for 180 days starting from  March 2$^{nd}$.
If the lockdown conditions are relaxed on May 17$^{th}$, the rate of transmission of infection is going to increase. In the relaxed lockdown, the $\alpha$ and $\beta$ values are assumed to jump by 20$\%$. The prediction for both cases, with a 4k increase in tests per day and saturation at 200k tests, is compared in FIG. \ref{India_ld_tad} and in FIG. \ref{India_ld1}, we plot the prediction band for the daily new cases considering two percent error in the estimation of rate of transmission of infection. We compare the effect of testing on the prediction in FIG. \ref{India_ld2}. Total, Active and extinct cases are plotted for the coming months if tests per day are increased by 4k and 8k per day after May 11$^{th}$ and saturated at 200k and 300k, respectively, taking into account the relaxation of lockdown after May 17$^{th}$. 
\begin{figure}[t]
        \centering
         \includegraphics[width=\linewidth]{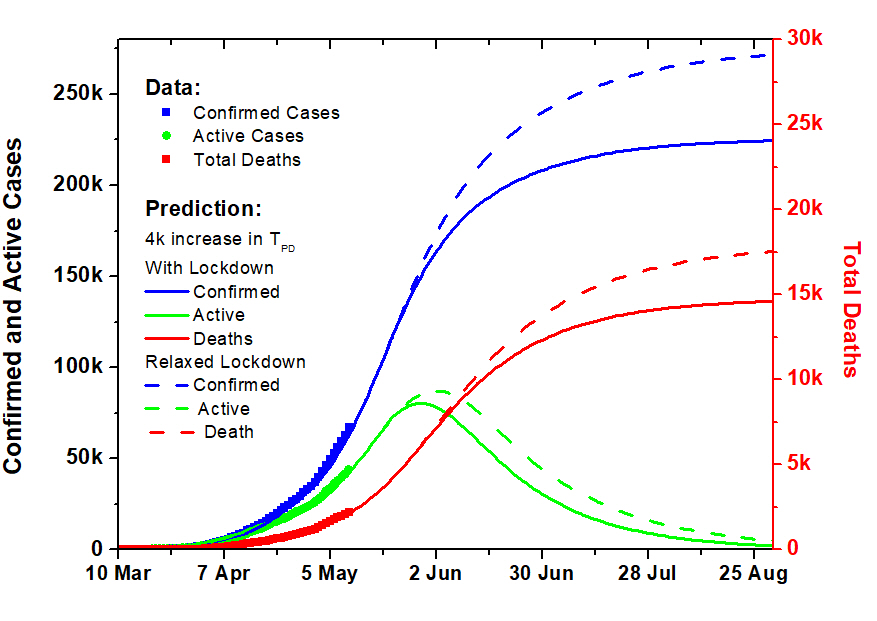}\par\caption{Comparison of the Model prediction for India for lockdown and relaxed lockdown conditions after May 17$^{th}$ with 4K increase in tests per day after May 11$^{th}$ and saturating at 200k.}\label{India_ld_tad}
            \end{figure}
   \begin{figure}[H]
        \centering
         \includegraphics[width=\linewidth]{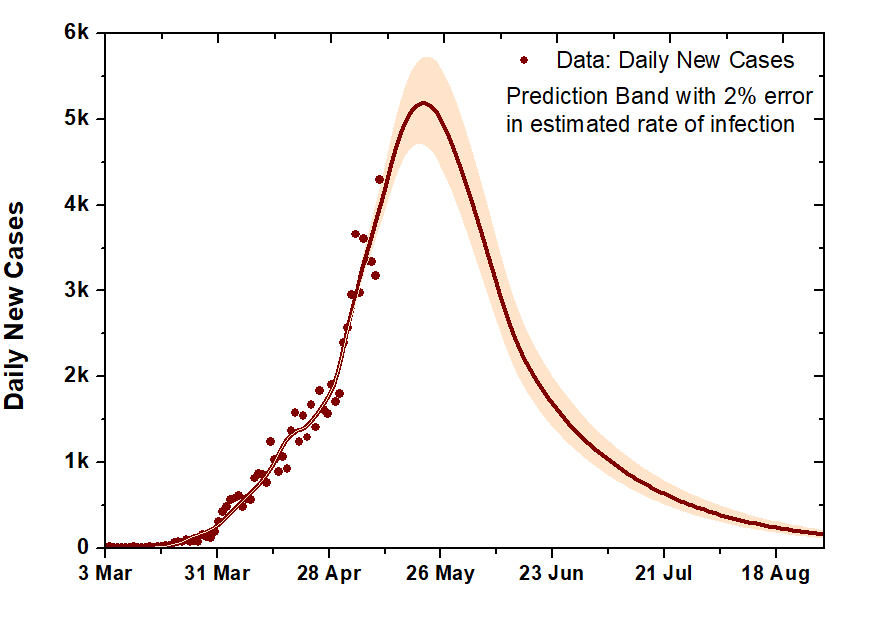}\par\caption{Comparison of the daily new cases prediction for India for lockdown conditions with 4K increase in tests per day by after May 11$^{th}$ and saturating at 200K. Shaded area shows the prediction band for 2 percent error in $\alpha_3$, $\beta_3$ estimation.}\label{India_ld1}
        \end{figure}
    \begin{figure}[h]
        \centering
         \includegraphics[width=\linewidth]{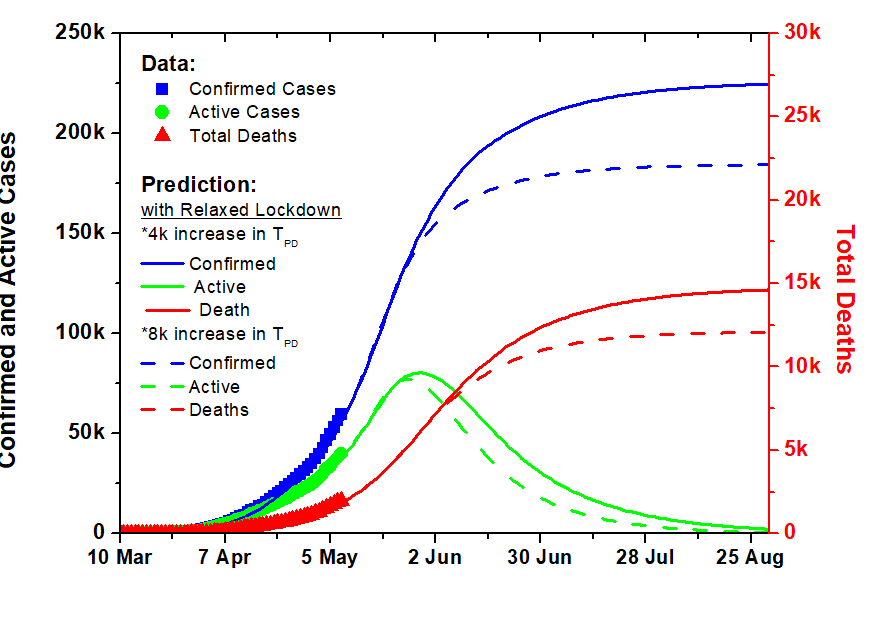}\par\caption{Comparison of the Model prediction for Confirmed, Active and total Deaths Cases for India. After May 11$^{th}$  tests are increased by 4k and 8k daily and saturated at 200K and 300K respectively, with lockdown relaxed after May 17$^{th}$.}\label{India_ld2}
    \end{figure}
\subsection{Predictions for USA}
The recovery rate of the $H$ category is found to be slow compared to South Korea or Germany, which may be attributed to either incorrect reporting of the Active cases or the testing of  serious cases only and longer recovery time in hospitals compared to quarantined with mild symptoms. The prediction for the next 240 days, that is till the end of the year 2020, is plotted in FIG. \ref{USA_ld_tad} for 10k test per day and saturated at 1 million tests per day. The recovery rate $\sigma$ is taken improved by 25 $\%$ after May 4$^{th}$ and mortality rate is taken improved in steps from initial 6E-3 to 4E-3 to 2.5E-3 after April 9$^{th}$ and April 19$^{th}$. The time evolution of the totally unknown and undetected part of the infected for USA is plotted in FIG. \ref{USA3}.
%The increase in the infection rate  if the lockdown  is relaxed is assumed  as 35 $\%$  from the current value. 
%It is   projected in   FIG.\ref{USA_ld_tad},  how fast the  
%disease can be contained  with an  increase by 20K  tests per day even in the scenario of  relaxed lockdown.
The daily new positive cases data and the prediction are plotted in FIG. \ref{USA_ld_dnc}.
%If lockdown is relaxed and due to that if there is 35$\%$ increase in the rate of transmission of infection than the current value then how fast the disease can be contained for 20k increase in tests per day is also plotted in FIG.\ref{USA_ld_tad}. The daily new cases for the above three different situations is plotted in FIG. \ref{USA_ld_dnc}.
\begin{table}[h]
\caption{Parameters values for the Countries studied} % title of Table
\centering % used for centering table
\begin{tabular}{c c c c c c c c c c c} \\% centered columns (10 columns)
\hline %inserts double horizontal lines
{\bf Para.}    & {Germany} & {S. Korea} & {India} & {USA}   \\ [0.5ex] 
\hline % inserts single horizontal line
Population N & { 8.30E7} & { 5.10E7} & { 1.38E9} & { 3.31E8} \\
$T_{0}$    & {21Feb}       & {15Feb}       & {3Mar}       & {15Feb}       \\
$T_{LD}$   & { 22}       & { 21}       & { 21,5}       & { 45}       \\
$\alpha$ (bf. LD)  &{ 0.32}  & { 0.39}  & { 0.33}  & { 0.33}   \\
$\alpha$ (af. LD)  &{ 0.30,0.2}  & { 0.085} & { 0.18,0.23}  & { 0.13}  \\
$\beta$ (bf. LD)   &{ 0.29,0.11}  & { 0.22}  & { 0.1}  & { 0.26}   \\
$\beta$ (af. LD)   &{ 0.24}  & { 0.074}   & { 0.05,0.18} & { 0.20}  \\
$\gamma$ (bf. LD)  &{ 0.32}  & { 0.39}  & { 0.33}  & { 0.33}   \\
$\gamma$ (af. LD)  &{ 0.29}  & { 0.085} & { 0.18}  & { 0.13}  \\
$\delta$           &{9E-3}   & { 9E-3}  & {9E-3}   & {9E-3}   \\
$\xi_I$     & { 0.2}  & { 0.2}   & { 0.2}   & { 0.2}      \\
$\xi_P$     & { 0.072}  & { 0.072}   & { 0.072}   & { 0.072}      \\
$\mu_0$            &{ 2.03E-6} & { 2.74E-5} & {1.13E-6} & {5E-7} \\
$\nu_0$  &{ 0.05}     & { 0.05}     & { 0.05}     & { 0.05}     \\
%$\nu_1$ & { 2.29}   & { 1.96}   & { 1.71}   & { 1.69}   \\
$\nu_1$/$\mu_0$    & { 5.34}   & { 1.09}   & { 4.98}   & {3.36}   \\
$\omega$           & { 0.1}    & { 0.1}    & { 0.1}    & { 0.1}      \\
$\eta$             & { 0.07}   & { 0.07}   & { 0.07}   & { 0.07}     \\
$\sigma$           &{0.059}    & {0.034}   & {0.047}   & {0.013}        \\
$\tau$             &{ 3.5E-3}   & { 0.9E-3}   & { 5E-3,3E-3}    & { 5E-3}   \\
$t_R$              &{ 5}        & { 7}        & { 7}      & { 10}        \\
$t_D$              &{ 5}        & { 4}        & { 1}      & { 1}      \\
Ini. Infected  &{50} & {150} & {100} & {40} \\
%$E(0)N$ &{20} & {100} & {30} & {70} \\
%$I(0)N$ &{8} & {50} & {10} & {40} \\
%$H(0)N$ &{2} & {19} & {3} & {12} \\
\hline
\end{tabular}
\label{countries_para}
\end{table}
\begin{figure}[h]
        \centering
         \includegraphics[width=\linewidth]{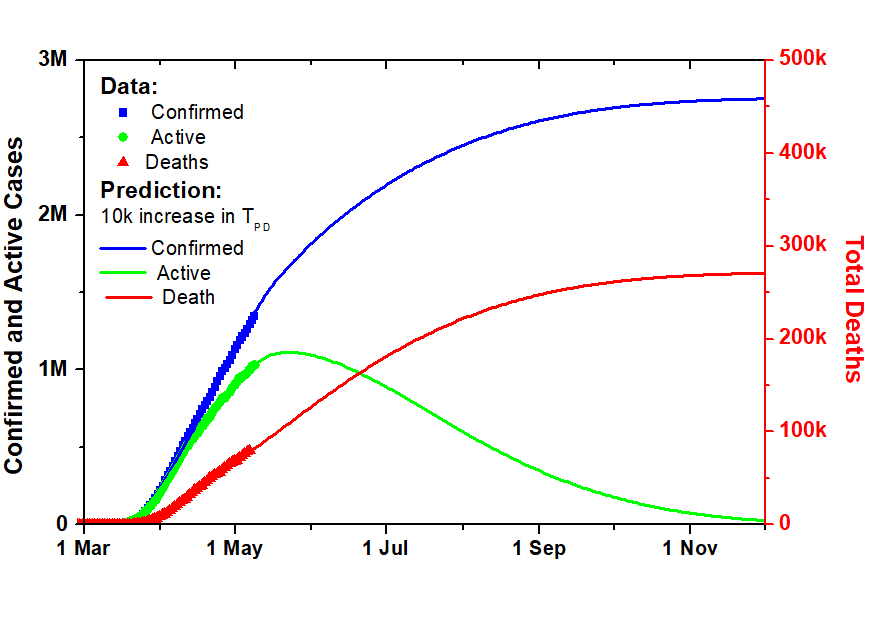}\par\caption{Comparison of the Model prediction for the USA if the tests increased by 10k daily and social distancing and lockdown conditions kept same.}\label{USA_ld_tad}
        \end{figure}
\begin{figure}[h]
    \includegraphics[width=\linewidth]{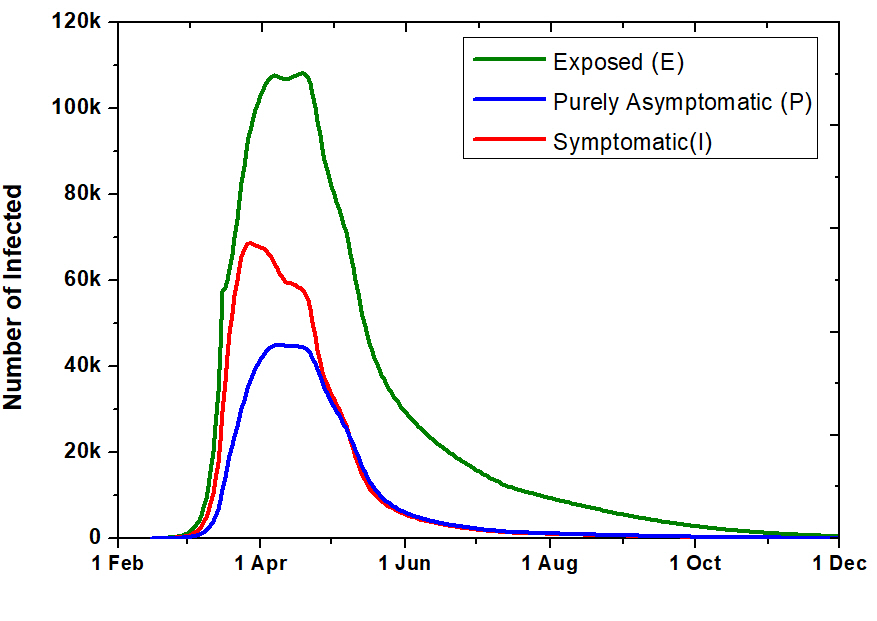}\par\caption{The evolution of the undetected number of Infected in Exposed (E), Symptomatic (I) and purely Asymptomatic (P) category for USA. }\label{USA3}
    \end{figure}

\begin{figure}[h]
        \centering
         \includegraphics[width=\linewidth]{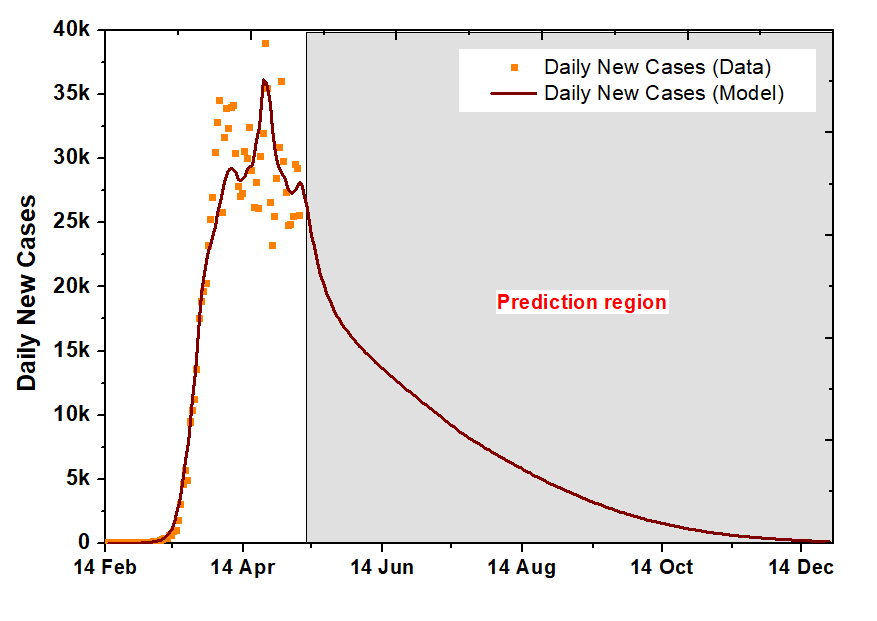}\par\caption{Comparison of the Model prediction for daily new cases for the USA with the increase of the 10k tests per day with lockdown. }\label{USA_ld_dnc}
    \end{figure}
  \end{section}
\begin{section}{Conclusion}
SIPHERD model is developed by considering purely Asymptomatic category of COVID-19 infected cases in addition to the Symptomatic, and the disease spread by the exposed. The effect of lockdown on the rates of transmission of infection and the influence of tests per day on detection rates has been incorporated in the model. The SIPHERD model is put for trial for the data of  South Korea and Germany, and with a limited number of days data, the  model is found to correctly predict the known evolution. The prediction for India suggests that even increasing  the rate of infection transmission by 20$\%$ due to relaxation of lockdown leads to around 50k increase in the total number of cases and 3k increase in total deaths. The prediction for the USA shows that in the absence of vaccine the infection can last long till the end of this year and number of deaths could be around 250k if lockdown and social distancing conditions remain the same. 
  
\subsection*{Acknowledgement}
AM would like to thank Dr. Shrikant Ambalkar, M.D for helpful discussions.
 \end{section}
\bibliographystyle{IEEEtran}
\bibliography{ref_covid}

\end{document}